\documentclass[10pt,a4paper,onecolumn, conference]{IEEEtran}

\newtheorem{thm}{Theorem}
\newtheorem{defn}{Definition}

\usepackage{graphicx}
\usepackage{amssymb,amsmath}
\usepackage{verbatim}

\begin{document}

%---------- Title ----------
\title{Myopic Coding in Multiple Relay Channels}

\author{\authorblockN{Lawrence Ong}
\authorblockA{Department of Electrical and Computer Engineering\\
National University of Singapore\\
Email: lawrence.ong@nus.edu.sg
%(version: \today)
}
\and
\authorblockN{Mehul Motani}
\authorblockA{Department of Electrical and Computer Engineering\\
National University of Singapore\\
Email: motani@nus.edu.sg}
}

\maketitle

%---------- Abstract ----------
\begin{abstract}
In this paper, we investigate achievable rates for data transmission from sources to sinks through multiple relay networks. We consider \emph{myopic coding}, a constrained communication strategy in which each node has only a local view of the network, meaning that nodes can only transmit to and decode from neighboring nodes.  We compare this with \emph{omniscient coding}, in which every node has a global view of the network and all nodes can cooperate. Using Gaussian channels as examples, we find that when the nodes transmit at low power, the rates achievable with two-hop myopic coding are as large as that under omniscient coding in a five-node multiple relay channel and close to that under omniscient coding in a six-node multiple relay channel.  These results suggest that we may do local coding and cooperation without compromising much on the transmission rate.  Practically, myopic coding schemes are more robust to topology changes because encoding and decoding at a node are not affected when there are changes at remote nodes. Furthermore, myopic coding mitigates the high computational complexity and large buffer/memory requirements of omniscient coding.
\end{abstract}

%===================== INTRODUCTION ====================
% 1 1/2 page

\section{Introduction}
\subsection{Multiple Relay Channels and Channel Constraints}
%Problems with current coding
The relay channel was first introduced by van der Muelen~\cite{meulen71} in his work on three terminal networks.  The capacity of a special class (known as the degraded relay channel) of the relay channel was found by Cover and El Gamal~\cite{covergamal79}.  In that paper, two coding strategies were proposed for the general relay channel, which were subsequently termed decode-forward strategy and compress-forward strategy.  Gupta and Kumar~\cite{guptakumar03} extended the relay channel to the multiple relay channel, where there is more than one relay node in the channel.  The decode-forward and the compress-forward strategies were extended to the multiple relay channel by Xie and Kumar~\cite{xiekumar03} and Kramer \emph{et al.}~\cite{kramergastpar04} respectively.  In these strategies, block Markov encoding (see \cite{covergamal79} for irregular block Markov and \cite{willemsmeulen83} for regular block Markov) is used.  In decoding, forward decoding~\cite{covergamal79} can be used for irregular Markov encoding and backward decoding~\cite{willemsmeulen85} or window decoding~\cite{kramergastpar03} can be used for regular block Markov encoding.  

\begin{figure}[t]
\centering
\includegraphics[width=8cm]{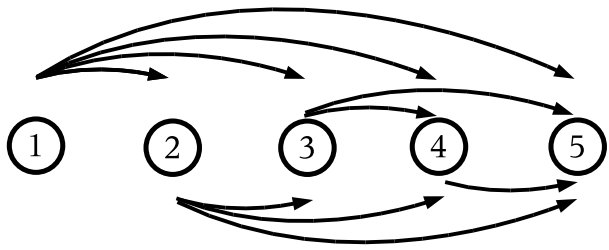}
\caption{Omniscient coding in a five-node Gaussian multiple relay channel.} \label{fig:5_node_mr_complete_view}
\end{figure}

Consider the five-node Gaussian multiple relay channel depicted in Fig.~\ref{fig:5_node_mr_complete_view}.  Using the decode-forward strategy, node 1 splits its power to send different messages to nodes 2-5 during each transmission.  In decoding, each node decodes messages from the transmissions of all nodes \emph{behind} it.  As the effect of all nodes' transmissions is being considered in the coding design, a node needs to be aware of the presence of all other nodes and to have knowledge of their codebooks.  We see that encoding and decoding can get complicated, e.g., more processing and buffering, as the network size grows.  We call this unconstrained communication on the multiple relay channel with a global view and complete cooperation {\em omniscient coding}.

The simplest approach to data transmission is for a node to communicate with only one node at a time. This leads naturally to multi-hop routing, in which each node sends data to the next node in the route and decodes data from the previous node in the route.  The transmissions of the other nodes are treated as noise.  We term this highly constrained communication \emph{point-to-point coding}.  

In this paper, we look at the compromise between omniscient coding and point-to-point coding.  We study how encoding and decoding are done when a node \emph{sees} only a few other nodes.  We term this constrained communication with a local view and limited cooperation {\em myopic coding}.  We determine achievable rates of multiple relay channels under myopic coding, using regular block Markov encoding and window decoding.  However, the encoding and decoding techniques differ from that found in the literature (in \cite{xiekumar03} and \cite{kramergastpar03}) as the nodes have limited view.  We note that point-to-point coding and omniscient coding are limiting cases of myopic coding.

\subsection{Practical Advantages of Myopic Coding}
\begin{figure}[t]
\centering
\includegraphics[width=8cm]{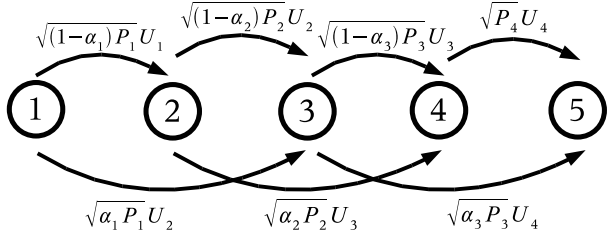}
\caption{Two-hop myopic coding in a five-node multiple relay channel.} \label{fig:5_node_mr_2_hop}
\end{figure}

%The benefits of using Myopic
Under omniscient coding, any topology change in the network,
for example node failure or mobility, requires reconfiguration of
coding and decoding at every node in the network.  This is due to the fact that a
node considers the transmission of all other nodes in its
encoding and decoding processes.  Myopic coding, however, does not suffer from this problem.  Using the five-node multiple relay channel as an example, Fig.~\ref{fig:5_node_mr_2_hop} depicts two-hop myopic coding, where a node only \emph{sees} nodes within
two hops away. Under this coding, when node 4 fails, no change is required at node 1, which is three hops away.

Besides being robust to topology changes, myopic coding
offers additional practical advantages over omniscient coding. Since a node only needs to
send signals to a few neighboring nodes, less computation is required at that node. Also, a node needs less memory for data buffering and codebook storage as decoding is done over a smaller decoding window size.

\subsection{Contributions}
We fist derive achievable rate regions for the multiple relay channel under two myopic coding constraints, namely one-hop coding and two-hop coding.  We use the concept of regular block Markov encoding to construct encoding methods for each node under one-hop coding and two-hop coding.  For decoding, we use the concept of window decoding, where the decoding of a message symbol is done over a few transmission blocks.

We compare achievable rates under myopic coding to that under omniscient coding.  We show that when nodes transmit at low power, the achievable rate region under two-hop coding is the same as (in a five-node multiple relay channel) and close to (in a six-node multiple relay channel) that achievable under omniscient coding. The achievable rate region under one-hop coding is close to that achievable under omniscient coding in a five-node channel but far below that under omniscient coding in a six-node channel.

We then extend the analysis to $k$-hop myopic coding, where $k > 2$ is a positive integer.  We construct encoding and decoding algorithms for $k$-hop coding and derive an achievable rate region. We also show that achievable rates under myopic coding are bounded away from zero even as the total number of nodes in the network grows large.

%=========================BACKGROUND THEORY==========================
% 1/2 oage

\section{Channel Model and Notations}

\begin{figure}[t]
\centering
\includegraphics[width=10cm]{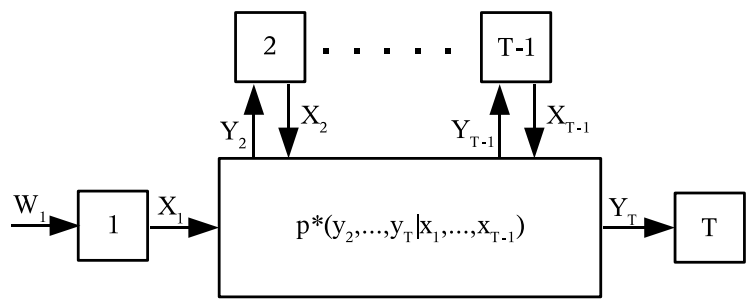}
\caption{A $T$-node multiple relay channel.} \label{fig:1DSensor}
\end{figure}

%----------------------Channel Model---------------------------
Fig.~\ref{fig:1DSensor} depicts a $T$-node multiple relay channel, with node 1 being the source node and node $T$ being the destination node.  Nodes 2 to $T-1$ are relay nodes. The message
$W$ is generated at node 1 and is to be sent to the
sink at node $T$.  A memoryless multiple relay channel can be
completely described by the channel distribution
\begin{equation}
p^*(y_2, y_3, \dotsc, y_T | x_1, x_2, \dotsc, x_{T-1})
\end{equation}
on $\mathcal{Y}_2 \times \mathcal{Y}_3 \times \dotsm \times \mathcal{Y}_T$, for each $(x_1, x_2, \dotsc, x_{T-1}) \in \mathcal{X}_1 \times \mathcal{X}_2 \times \dotsm \times \mathcal{X}_{T-1}$.  In this paper, we only consider memoryless channels, which means
\begin{equation}
p^*(\mathbf{y}_2^n, \mathbf{y}_3^n, \dotsc, \mathbf{y}_T^n | \mathbf{x}_1^n, \mathbf{x}_2^n, \dotsc, \mathbf{x}_{T-1}^n) = \prod_{i=1}^n p^*(y_{2,i}, y_{3,i}, \dotsc, y_{T,i} | x_{1,i}, x_{2,i}, \dotsc, x_{T-1,i})
\end{equation}
where $\mathbf{x}_j^n = (x_{j,1}, x_{j,2}, \dotsc, x_{j,n})$ is an ordered vector of $x_j$ of size $n$.

%--------------------------Definitions---------------------------
Standard terms, such as codebook, error probability, typical sequences, and achievable rates, are the same as those defined in \cite{coverthomas91}.  When the terms carry different meanings, they will be explicitly defined. We define myopic coding as follows.
\begin{defn}
$k$-hop myopic coding is defined as constrained communication among nodes in the multi-terminal network satisfying the following:
\begin{itemize}
    \item In encoding, a node can only transmit messages that it has decoded or compressed from the past $k$ blocks of received signal.
    \item A node can only store a decoded message in its memory over at most $k$ blocks.
    \item In decoding, a node can only decode/process one message using only $k$ blocks of received signal.
\end{itemize}
\end{defn}
We note that the notion of the ``view'' of a node, meaning how many other nodes a node can see, is embedded in the definition itself. This definition allows myopic coding to be easily extended to other types of channels, for instance, the broadcast multiple relay channel and the multiple access relay channel \cite{OngMotaniTR05}. Also, the rationale of myopic coding stems from the advantage of having less processing and less storage at a node.

 %==========================ACHIEVABLE RATES========================
% 2 pages

\section{Achievable Rates Under Different coding}
Let $\mathcal{R}$ be the set of all relay nodes, $\mathcal{R} = \{ 2, 3, \dotsc, T-1 \}$, and let
$\pi(\cdot)$ be a permutation on $\mathcal{R}$.  Define $\pi(1) =
1$, $\pi(T) = T$ and $\pi(i:t) = \{ \pi(i), \pi(i+1), \dotsc, \pi(t)
\}$.

%------------------------- One-hop View --------------------------
\subsection{One-Hop Myopic Coding}
Under one-hop coding, each node only sends signals to
the node \emph{in front} of it and decodes signals from the node
\emph{behind} it. We assume perfect echo cancellation, which means that a node is able to cancel the effect of its own transmission in its received signals.  Using non-constructive coding \cite{shannon48}, node $t$ can receive information up to the following rate.
\begin{equation}
R_t \leq \max I(X_{t-1};Y_t|X_t)
\end{equation}
for $t \in \{ 2, \dotsc, n\}$ and $X_T=0$.  The maximization is over
the distribution $p(x_1)p(x_2)\dotsm p(x_{T-1})$. Since all
information must pass through all nodes in order to reach the
destination, the overall rate is constrained by
\begin{equation}
R \leq \max_{\pi(\cdot)} \max_{p(\cdot)} \min_{t \in \{ 2, \dotsc, T\}} I(X_{t-1};Y_t|X_t).
\end{equation}

%----------------------- Two-Hop View -------------------------------
\subsection{Two-Hop Myopic Coding}
Instead of just transmitting to one node in front, a node might want to help the node in front to transmit to the node that is two hops away. The nodes can do that in two-hop myopic coding.  Equivalently, in block $i$, a node transmits data that it has decoded in blocks $i-1$ and
$i-2$. In decoding, it decodes one message using only two blocks of received signals. We consider $B+T-2$ transmission blocks, each of $n$
uses of the channel. A sequence of independent $B$ indices, $w(b)
\in \{ 1, 2, \dotsc, 2^{nR} \}$, $b = 1, 2, \dotsc, B$ will be sent
over $n(B+T-2)$ uses of the channel.  As $B \rightarrow \infty$, the
rate $RnB/n(B+T-2) \rightarrow R$ for any $n$.

\subsubsection{Codebook Generation}
In this section, we describe how codebooks at each node are generated.

\begin{itemize}
    \item First, fix the probability distribution
    \begin{equation}
    p(u_1,u_2,\dotsc,u_{T-1},x_1,x_2,\dotsc,x_{T-1})
    = p(u_1)p(u_2)\dotsm p(u_{T-1}) p(x_1|u_1,u_2) p(x_2|u_2,u_3)
    p(x_{T-1}|u_{T-1}),
    \end{equation}
    for each $u_i \in \mathcal{U}_i$.
    \item For each $t \in \{1,\dotsc,T-1\}$, generate $2^{nR}$ independent and identically distributed
    (i.i.d.) $n$-sequences in $\mathcal{U}_t^n$, each drawn
    according to $p(\mathbf{u}_{t}) = \prod_{i=1}^n p(u_{t,i})$.  Index
    them as $\mathbf{u}_{t}(w_{t})$, $w_{t} \in \{1,\dotsc, 2^{nR}\}$.
    \item Define $\mathbf{x}_{T-1}(w_{T-1}) = \mathbf{u}_{T-1}(w_{T-1})$.
    \item For each $t \in \{1,\dotsc,T-2\}$, define a deterministic function
    that maps $(\mathbf{u}_t,\mathbf{u}_{t+1})$ to $\mathbf{x}_t$:
    \begin{equation}
    \mathbf{x}_t(w_t,w_{t+1}) =
    f_t \big( \mathbf{u}_t(w_t),\mathbf{u}_{t+1}(w_{t+1}) \big).
    \end{equation}
    \item Steps 2 to 4 are repeated to generate a new independent set
    of codebooks.  These two codebooks are used in alternate transmission blocks.
\end{itemize}

We see that in each transmission block, node $t$, $t \in \{1,\dotsc,T-2\}$,
sends messages of two blocks $w_t$ (new data) and $w_{t+1}$ (old
data). In the same block, node $t+1$ sends messages $w_{t+1}$ and
$w_{t+2}$. Note that a node cooperates with the node {\em in front}
by repeating the transmission $w_{t+1}$.  Subscript $t$ represents new data that is being sent by node $t$.

\subsubsection{Encoding}
\begin{figure}[t]
    \centering
        \includegraphics[width=12cm]{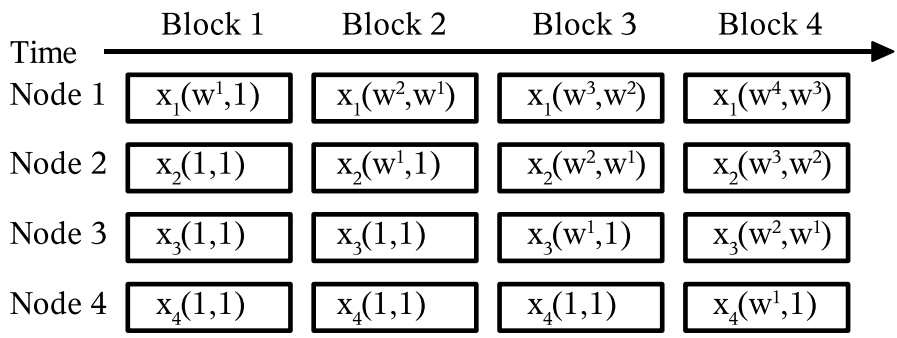}
    \caption{A two-hop encoding strategy.}
    \label{fig:two_hop_encoding}
\end{figure}

Fig.~\ref{fig:two_hop_encoding} shows the encoding process for two-hop coding.  The encoding steps are as follows:
\begin{itemize}
    \item In the beginning of block 1, the information source emits the first source letter $w^1$.
    Here, we use superscript to indicate the time index of the source
    letter.  That is, the source emits $w^1, w^2, \dotsc, w^b$ at the
    beginning of block $1, 2, \dotsc, b$ respectively.  Note that
    there is no new information after block $B$.  We define $w^{B+1} = w^{B+2} = \dotsm = w^{B+T-2}
    = 1$.
    \item In block 1, node 1 transmits $\mathbf{x}_1(w^1,w^0)$. Since
    the rest of the nodes have not received any information, they
    send the dummy letter $\mathbf{x}_i(w^{2-i},w^{1-i})$, $i \in \{2,\dotsc,T-1\}$.  We define $w^b=1$, for $b \leq 0$.
    \item At the end of block 1, assuming that node 2 correctly decodes the first signal
    $w^1$, it transmits $\mathbf{x}_2(w^1,1)$.
    \item Generalizing, in block $b \in \{1,\dotsc,B+T-2\}$, node $t$, $t \in \{1,\dotsc,T-1\}$, would have decoded data $(w^1, w^2, \dotsc, w^{b-t+1})$ and it sends $\mathbf{x}_t(w^{b-t+1},w^{b-t})$.
\end{itemize}

\begin{figure*}[t]
\centering
  \includegraphics[width=12cm]{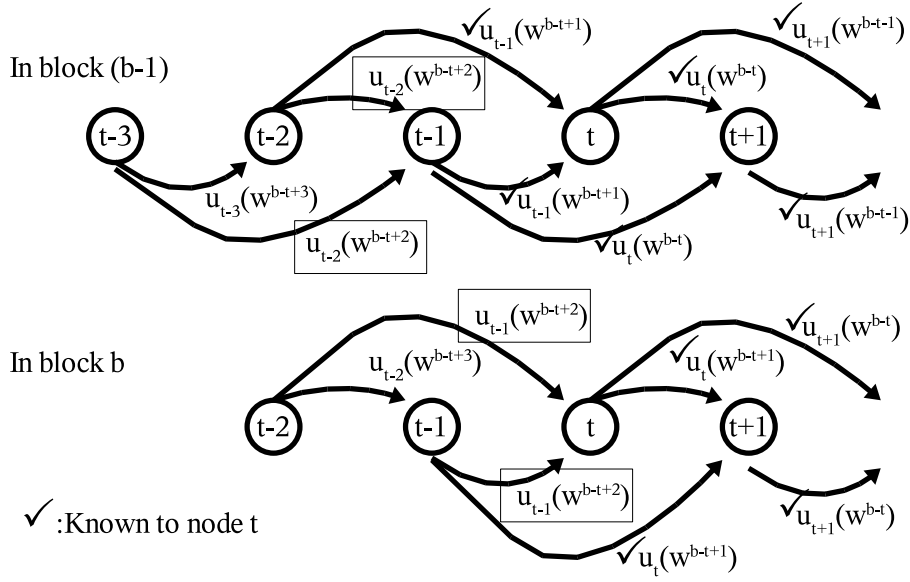}\\
  \caption{Decoding at node $t$ of message $w^{b-t+2}$.}\label{fig:decoding_ex}
\end{figure*}
\subsubsection{Decoding and Achievable Rates}
All nodes except node 2 decode one message over two blocks of the received signal. As depicted in Fig.~\ref{fig:decoding_ex}, node $t$ decodes the message $w^{b-t+2}$ over blocks $(b-1)$ and $b$.  
%From block $b$, node $t$ can decode $w^{b-t+2}$ up to $I(U_{t-1};Y_t|U_t,U_{t+1})$.  
%From block $b-1$, node $t$ can decode $w^{b-t+2}$ up to rate $I(U_{t-2};Y_t|U_{t-1}, U_t,U_{t+1})$.  
%Since the same message is decoded over both blocks, the rate at which the message $W$ is decodable at node $t$ is $I(U_{t-2},U_{t-1};Y_t|U_t,U_{t+1})$.
It can be shown that the rate at which the message $W$ is decodable at node $t$ is $I(U_{t-2},U_{t-1};Y_t|U_t,U_{t+1})$.  It can be shown that the probability of error can indeed be made as small as desired if the rate constraint above is satisfied.  The proofs, given in \cite{OngMotaniTR05}, are omitted due to space limitations.

\begin{thm}
In a $T$-node memoryless multiple relay channel, under two-hop coding, the following rate is achievable,
\begin{equation}
R \leq \max_{\pi(\cdot)} \max_{p(\cdot)} \min_{t \in \{2, \dotsc, T\}} I(U_{t-2},U_{t-1};Y_t|U_t,U_{t+1}),
\end{equation}
where $U_0 = U_T = U_{T+1} = 0$ and the maximization is taken over all joint distributions of the form
\begin{subequations}
\begin{align}
    &p(x_1, x_2 \dotsc, x_{T-1}, u_1, u_2 \dotsc, u_{T-1}, y_2, y_3 \dotsc, y_T)\nonumber\\
    & = p(u_1)p(u_2)\dotsm p(u_{T-1}) p(x_1|u_1,u_2) p(x_2|u_2,u_3) \dotsm \nonumber\\
    & \quad p(x_{T-1}|u_{T-1})p^*(y_2, \dotsc, y_T | x_1, \dotsc, x_{T-1}).
\end{align}
\end{subequations}
\end{thm}

%------------------------- omniscient ---------------------------
\subsection{Omniscient Coding}
Omniscient coding was considered by Xie and Kumar~\cite{xiekumar03}.  Using the decode-forward strategy, they showed that the following rate is achievable,
\begin{equation}
R \leq \max_{\pi(\cdot)} \max_{p(\cdot)} \min_{
1\leq t \leq T-1} I(X_{\pi(1:t)} ; Y_{\pi(t+1)} | X_{\pi(t+1;T-1)}).
\end{equation}
The first maximization allows us to arrange the order of the relay
nodes in which the data flows through them.  The second maximization
is over all possible distributions $p(x_1,x_2,\dotsc,x_{T-1})$.  The
minimization is on the rate at which each relay node receives.  This is because each node needs to fully decode every message.

\begin{figure}[t]
\centering
\includegraphics[width=10cm]{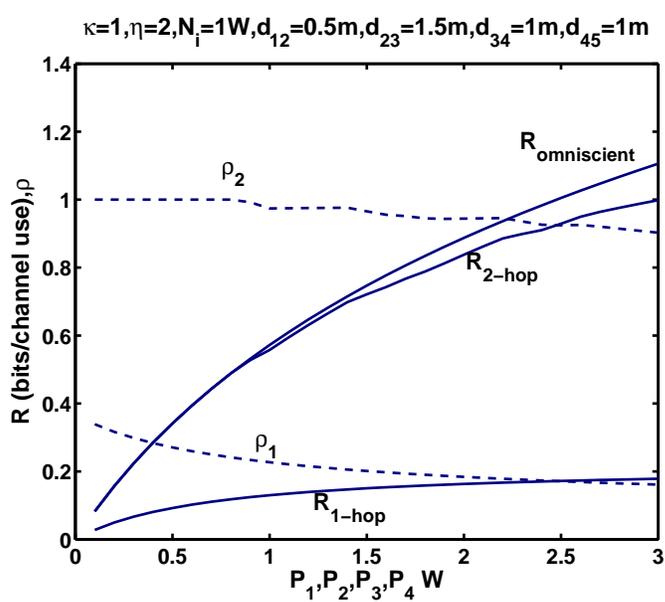}
\caption{Achievable rates under different coding constraints in a five-node
multiple relay channel.} \label{fig:rate_power_5_node}
\end{figure}

%======================== PERFORMANCE COMPARISON ====================
% 1 page

\section{Performance Comparison on Gaussian Channels}

For the purpose of comparison, we study the performance of the different schemes on Gaussian channels of the form,
\begin{equation}\label{eq:gaussian_rate}
Y_t = \sum_{\substack{i=1\\i\neq t}}^{T-1} \sqrt{\kappa d_{it}^{-\eta}}X_i
+ Z_t, \quad\quad t = 2, 3, \dotsc, T
\end{equation}where $X_i$ is a random variable with power constraint $E[X_i^2] \leq P_i$ and $Z_t$ is the receiver noise, which is a zero mean Gaussian random variable with variance $N_t$.  We use the standard path loss model for signal propagation, where $d_{it}$ is the distance between node
$i$ and node $t$, $\kappa$ is a positive constant, $\eta$ is the path loss exponent, and
$\eta \geq 2$ with equality for free space transmission. Also, we consider networks where the nodes are arranged in  a straight line.

\begin{figure}[t]
\centering
\includegraphics[width=10cm]{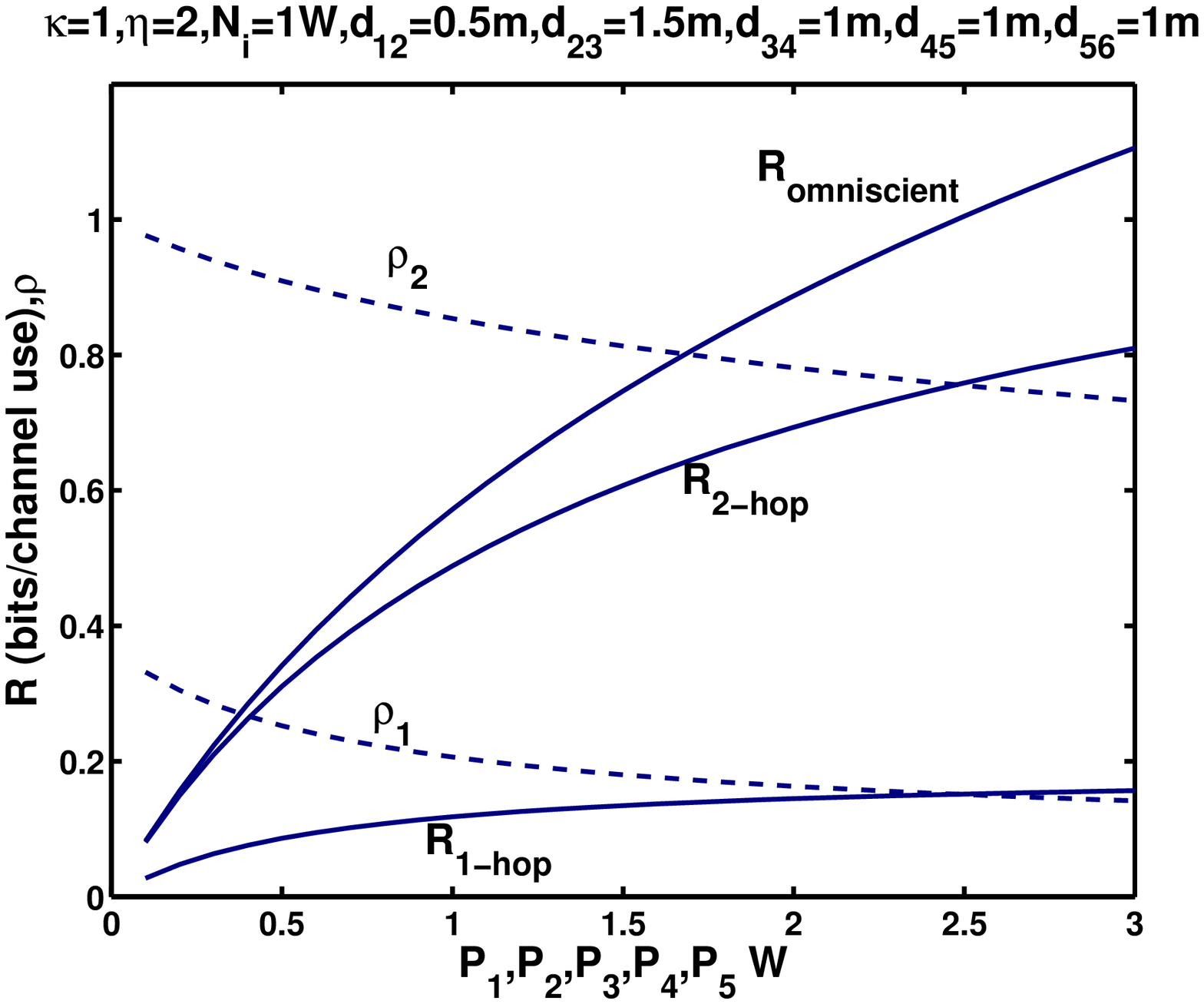}
\caption{Achievable rates under different coding constraints in a six-node
multiple relay channel.} \label{fig:rate_power_6_node}
\end{figure}

Figures~\ref{fig:rate_power_5_node} and \ref{fig:rate_power_6_node} show achievable rates under one-hop coding, two-hop coding, and omniscient coding in a five-node and a six-node Gaussian multiple relay channels respectively. 
We make the following observations.

\begin{itemize}
\item
As expected, the achievable rates under myopic coding are not more than that under omniscient coding. 
However, at low SNR, we see that myopic coding is close to omniscient coding.

\item 
We note that achievable rates increase significantly from one-hop to two-hop coding.  This suggests that for a multiple relay channel with many nodes, myopic coding with ''short'' view is sufficient.

\item
We define $\rho_i = R_{\text{myopic}}/R_{\text{omniscient}}$ where $i=1,2$ for one-hop and two-hop coding respectively.  When the number of nodes increases, $\rho_1$ and $\rho_2$ decrease.  This is because more nodes are ignored in myopic coding when the channel size gets larger.  However, in a six-node channel, two-hop coding can still achieve $\rho_2 > 0.8$ for transmit SNR smaller than 1dB.

\end{itemize}

%======================= EXTENSION TO K-HOP VIEW ========================   
% 1/2 page

\section{Extending to $k$-hop coding}
Now, we generalize two-hop coding to $k$-hop coding where $k \in \{1,\dotsc,T-1\}$. The proof is given in \cite{OngMotaniTR05} but omitted here. 

\begin{thm}
In a $T$-node memoryless multiple relay channel, using $k$-hop coding, the following rate is achievable.
\begin{equation}
R \leq \max_{\pi(\cdot)} \max_{p(\cdot)} \min_{t \in \{2, \dotsc, T\}} I(U_{t-k},\dotsc,U_{t-1};Y_t|U_t,\dotsc ,U_{t+k-1})
\end{equation}
where $U_{2-k} = U_{3-k} = \dotsm = U_0 = U_T = U_{T+1} = \dotsm = U_{T+k-1} = 0$ and the maximization is taken over all joint distributions of the form
\begin{subequations}
\begin{align}
		p(x_1, x_2 \dotsc, x_{T-1}, u_1, u_2 \dotsc, u_{T-1}, y_2, y_3 \dotsc, y_T)
    & = \prod_{i=1}^{T-1}p(u_i) \prod_{i=1}^{k}p(x_{T-i}|u_{T-i},\dotsc,u_{T-1})
    \prod_{i=i}^{T-k-1} p(x_i|u_i,\dotsc,u_{i+k-1}) \nonumber\\
    & \quad \times p^*(y_2, \dotsc, y_T | x_1, \dotsc, x_{T-1}).
\end{align}
\end{subequations}
\end{thm}

%======================= LARGE NETWORK ============================
%===================================================================
\section{Myopic Coding in Large Networks}
Since the transmission beyond the view of a node is treated as noise, one concern with myopic coding is whether the rate vanishes as the number of nodes grows. We analyze two-hop coding in a $T$-node multiple relay channel, assuming that the nodes are equally spaced at 1m apart and transmit at power $P'$.  Considering the reception of node $t$, the signal power is given by
\begin{subequations}
\begin{align}
P_{sig}(t) & = \left( \sqrt{ 3^{-\eta}\alpha_{t-3}P} + \sqrt{2^{-\eta}(1-\alpha_{t-2})P} \right)^2\nonumber\\
& \quad + \left( \sqrt{ 2^{-\eta}\alpha_{t-2}P} + \sqrt{ 1^{-\eta}(1-\alpha_{t-1})P}\right)^2 > 0.
\end{align}
\end{subequations}
where $P = \kappa P'$.

The noise power is $P_{noise}(t) = N_t < \infty$ and the interference power is given by
\begin{subequations}
\begin{align}
\frac{P_{int}(t)}{P} & = 3^{-\eta}\alpha_{t-3} + \sum_{k=4}^{t-1} \frac{1}{k^{\eta}} + 1^{-\eta}\alpha_{t+1} + \sum_{k=2}^{T-t-1} \frac{1}{k^{\eta}} \nonumber\\
& \quad + 2 \sum_{k=3}^{t-2} \sqrt{ \frac{(1-\alpha_{t-k})\alpha_{t-(k+1)}}{k^{\eta}(k+1)^{\eta}} }+ 2 \sum_{k=1}^{T-t-3} \sqrt{ \frac{\alpha_{t+k}(1-\alpha_{t+k+1})} {k^{\eta}(k+1)^{\eta}} } \label{eq:14a}.
\end{align}
\end{subequations}
Noting that $0\le \alpha_t\le 1, \forall t$ and simplifying \eqref{eq:14a}, we get
\begin{equation}
\frac{P_{int}(t)}{P} < 6 \sum_{k=1}^{T} \frac{1}{k^{\eta}} < 6 \zeta(\eta).
\end{equation}
%Simplifying, we get
%\begin{subequations}
%\begin{align}
%\frac{P_{int}(t)}{P} & < \sum_{k=3}^{t-1} \frac{1}{k^{\eta}} + \sum_{k=1}^{T-t-1} \frac{1}{k^{\eta}} + 2\sum_{k=3}^{t-2} \frac{1}{k^{\eta}} + 2\sum_{k=1}^{T-t-3} \frac{1}{k^{\eta}}\\
%& < 6 \sum_{k=1}^{T} \frac{1}{k^{\eta}} < 6 \zeta(\eta).
%\end{align}
%\end{subequations}
Here $\zeta(\eta) = \sum_{k=1}^\infty \frac{1}{k^\eta}$ is the Riemann zeta function, which is a decreasing function of $\eta$. Since, the path loss exponent is always greater than 2, $P_{int}(t) < 6 \zeta(2)P  = \pi^2P$. Hence, we can always find set of $\{\alpha_1, \dotsc, \alpha_{T-2}\}$ such that the reception rate at every node $t$, $\forall t \in \{2, 3, \dotsc, T \}$, is
\begin{equation}
R_t = \frac{1}{2} \log \left[ 1 + \frac{P_{sig}(t)}{P_{int}(t) + N_t} \right] > 0.
\end{equation}

When more nodes are included in the ``view'' in myopic coding, $P_{sig}$ increases and $P_{int}$ decreases. In general, assuming that the nodes are roughly equally spaced, the achievable rates under $k$-hop myopic coding ($k \geq 2$) are bounded away from zero even as the network size grows to infinity.

%=================== CONCLUSION ===================
% 1/2 page

\section{Conclusion}
In this paper, we compare myopic coding, i.e., local view and limited cooperation, and omniscient coding, i.e., global view and complete cooperation, on multiple relay channels.  We compute achievable rates for myopic coding in a $T$-node multiple relay channel, using regular block Markov encoding and window decoding.
Our experiments with five-node and six-node relay channels showed a significant rate improvement from one-hop to two-hop coding and that two-hop coding can be as good as omniscient coding.  These observations demonstrate the benefits of local cooperation and that only a small fraction of the nodes need to cooperate.
This suggests that local coding design may be good enough without compromising rate.  

%We show that, at low transmit power, achievable rates under two-hop coding can be as large as that under omniscient coding in a five-node multiple relay channel, and close to that under omniscient coding in a six-node multiple relay channel. We derive achievable rates under $k$-hop coding ($2 \leq k  \leq T-1$) in the $T$-node multiple relay channel. We show that achievable rates under myopic coding are bounded away from zero even if the number of nodes in the network grows.

%---------- Bibliography -----------
\bibliography{bib}

\end{document}